\documentclass[aip,jcp,reprint,superscriptaddress,floatfix,longbibliography,amsfonts,amssymb,amsmath]{revtex4-1}

\usepackage{graphicx}
\usepackage{dcolumn}
\usepackage{bm}
\usepackage{hyperref}
\usepackage{color}

\newcommand{\blue}[1]{{#1}}

\begin{document}

\title{Large permeabilities of hourglass nanopores: from hydrodynamics to single file transport}
\date{\today}

\author{Simon Gravelle}
\affiliation{Institut Lumi{\`e}re Mati\`ere, Universit\'e Claude Bernard Lyon 1-CNRS, UMR 5306, Universit\'e de Lyon, F-69622 Villeurbanne cedex, France}
\author{Laurent Joly}
\affiliation{Institut Lumi{\`e}re Mati\`ere, Universit\'e Claude Bernard Lyon 1-CNRS, UMR 5306, Universit\'e de Lyon, F-69622 Villeurbanne cedex, France}
\author{Christophe Ybert}
\affiliation{Institut Lumi{\`e}re Mati\`ere, Universit\'e Claude Bernard Lyon 1-CNRS, UMR 5306, Universit\'e de Lyon, F-69622 Villeurbanne cedex, France}
\author{Lyd\'eric Bocquet}
\altaffiliation[Present adress: ]{Laboratoire de Physique Statistique, UMR CNRS 8550, Ecole Normale Sup\'erieure, 24 rue Lhomond, 75231 Paris Cedex 05, France}
\affiliation{Institut Lumi{\`e}re Mati\`ere, Universit\'e Claude Bernard Lyon 1-CNRS, UMR 5306, Universit\'e de Lyon, F-69622 Villeurbanne cedex, France}

\begin{abstract}
In fluid transport across nanopores,
there is a fundamental dissipation that arises from the connection between the pore and the macroscopic reservoirs. This entrance effect can hinder the whole transport in certain situations, for short pores and/or highly slipping channels. 
In this paper, we explore the hydrodynamic permeability of hourglass shape nanopores using molecular dynamics (MD) simulations, with the central pore size ranging from several nanometers down to a few Angstr\"oms. Surprisingly, 
we find a very good agreement between MD results and continuum  hydrodynamic predictions, even for the smallest systems undergoing single file transport of water. An optimum of permeability is found for an opening angle around 5$\,^\circ$, in agreement with continuum predictions, yielding a permeability five times larger than for a straight nanotube. Moreover, we find that the permeability of hourglass shape nanopores is even larger than single nanopores pierced in a molecular thin graphene sheet.  This suggests that designing the geometry of nanopores may help considerably increasing the macroscopic permeability of membranes. 
\end{abstract}



\maketitle

\section{Introduction}

The study of flow through nanoscale channels has generated a large activity over the last decades \cite{Eijkel2005,Bocquet2010}. 
On the experimental side, transport across biological channels, as well as through artificial nanopores such as solid-state nanopores, nano-slits, or carbon/boron-nitride nanotubes, have been thoroughly investigated \cite{Bocquet2010,Dekker2007}. This has unveiled a number of new behaviors, such as the fast flows through carbon nanotube membranes \cite{Holt2006a,Majumder2005,Whitby2007}, non-linear electrokinetic transport \cite{Karnik2007,Siwy2002,Bocquet2010}, or large osmotic transport in boron nitride tubes \cite{Siria2013}. 
On the theoretical side, the question of fluidic transport at the nanoscale has generated a considerable activity, involving in particular molecular dynamics (MD) simulations of flows in nanochannels. 
Since the pioneering work of Hummer \cite{Hummer2001},  nanochannels made of carbon nanostructures have been specifically explored \cite{Park2014}. 
This has led to the recent highlighting of the special properties of flows inside carbon nanotubes \cite{Park2014,Joseph2008,Falk2010,Goldsmith2009}, and more recently through nanopores pierced in graphene \cite{Cohen-Tanugi2012,Suk2013}. 


Beyond the sole transport properties in nanoconfinement, the question of entrance effects is of particular importance for pores with nanoscale dimensions. This  was discussed exhaustively in the physiology literature since the work of Hille and Hall \cite{Hille1968,Hall1975}, in the context of the entrance contribution to the electric resistance in ionic channels. However, hydrodynamic entrance effects for the flow resistance into a pore were already discussed a century ago by Sampson who calculated the flow  resistance across a circular pore in an infinitely thin membrane, within the framework of continuum hydrodynamics \cite{Sampson1891}. It was then generalized to a circular cylinder with finite length \cite{Weissberg1962, Weinbaums}. 


Hydrodynamic entrance effects involve a supplementary viscous dissipation, which  is expected to be dominant and limitating in systems of 
small length as compared to their radius, and/or systems with low friction (high slippage) at their walls. 
In this context, we showed in a recent contribution \cite{Gravelle2013}, using finite element (FE)  calculations and theoretical estimates, that the hourglass geometry with a small cone angle reduces considerably the entrance dissipation and accordingly increases the permeability of the nanopore. Interestingly, the corresponding geometry matches quite nicely the shape of aquaporins, a biological water filter known to exhibit large flow rates \cite{Murata2000,Sui2001,Borgnia1999,Lecture2003}. 
Now, biological nanopores, such as aquaporins, usually involve a single-file transport of water inside the core of the channel and
continuum hydrodynamics is accordingly expected to fail in the core of such sub-nanometric systems \cite{Thomas2009,Schoch2008,Bocquet2010}. This raises therefore the question of entrance effects at the interface between macroscopic reservoirs and single-file systems. How do entrance effects behave in such geometries? Does the hourglass shape retains its enhanced transport behavior, as suggested by continuum calculations? 

In the present paper we address these questions by exploring water transport across artificial hourglass-like channels using MD simulations. We show that -- in a quite unexpected way -- the results for transport in molecular pores compare \emph{quantitatively} with continuum hydrodynamics predictions. In particular, an optimized permeability is still found for a small cone angle around 5$\,^\circ$. We then  compare the hydrodynamic permeability across hourglass shaped pores to that of pierced graphene and show that the hourglass shape is significantly more efficient in terms of water transport.



\section{Hydrodynamic resistance, entrance effects and Sampson formula}

In the framework of weakly out-of-equilibrium systems, the hydrodynamic resistance $R_{\rm hyd}$ characterizes flow transport across a given structure, and is defined as the ratio between the pressure drop $\Delta P$ and the corresponding flow rate $Q$ between two reservoirs connected by the structure:
\begin{equation}
R_{\rm hyd} = \frac{\Delta P}{Q} .
\end{equation}
For a cylindrical channel with radius $a$ and length $L$,  the ``inner'' hydrodynamic resistance, relating the pressure drop along the channel to the flow rate, is given within continuum hydrodynamics 
by the well-known Poiseuille law \cite{Happel}: 
\begin{equation}
R_{\rm in}^{\rm no-slip} = \frac{8\eta L}{\pi a^4} ,
\end{equation}
where $\eta$ is the liquid viscosity, and a no-slip boundary condition (BC) has been assumed at the channel wall. 
In the case of liquid/solid slip at the pore wall, dissipation and accordingly the hydrodynamic resistance are reduced \cite{Joly2011}:
\begin{equation}
\label{poiseuille_b}
R_{\rm in}^{\rm slip} = \frac{R_{\rm in}^{\rm no-slip}}{1+\frac{4b}{a}} ,
\end{equation}
where $b$ is the so-called slip length \cite{Bocquet2010}. 


Beside this \textit{inner} resistance, entering the pore generates a supplementary hydrodynamic resistance, originating from the bending of the streamlines towards the pore. This question was first discussed by Sampson in an article in 1891 \cite{Sampson1891}, where he calculated the velocity profile of a liquid flowing through an infinitely thin  membrane pierced with a circular hole, using continuum hydrodynamics. 
He obtained the relationship between the flow rate $Q$ and the pressure drop $\Delta p$ as 
\begin{equation}
Q = \frac{a^3}{C \eta} \Delta p ,
\label{sampson}
\end{equation}
where $a$ is the hole radius and $C = 3$. Later, it was shown that Sampson's formula could describe \emph{quantitatively} the access pressure drop (i.e. the excess pressure drop due to the entrances) through a cylindrical pore,
assuming the standard hydrodynamic no-slip BC of the liquid at the pore walls \cite{Weissberg1962}.
Therefore the  access resistance to a cylindrical pore is given by a formula \textit{\`{a} la} Sampson, as
\begin{equation}
\label{eq:sampson}
R_{\rm out} = \frac{\Delta p_{\rm out}}{Q} = \frac{C \eta}{a^3},
\end{equation}
with $C$ a numerical constant, $C \approx 3$. 

Gathering contributions, the total hydrodynamic resistance of a cylindrical pore with finite length is given as:
\begin{equation}
R_{\rm tot} = R_{\rm in} + R_{\rm out} = \frac{8\eta L}{ \pi a^4} \times \frac{1}{ 1+\frac{4b}{a}}+ \frac{C \eta}{a^3} .
\end{equation}

In the no-slip case, the entrance effect becomes dominant when the aspect ratio $L/a$ is below a threshold
value $[L/a]_c = C\pi/8\sim 1$. 
In the large slip limit $b\rightarrow \infty$, $R_{\rm in}$ and $R_{\rm out}$ share the same scaling with the pore radius. Therefore, \emph{independently of the pore radius}, the access resistance become the limiting factor for a
critical pore length $L_{\rm c}$ such that 
$L_{\rm c} =  \frac{\pi C}{2} b$.
For $L \ll L_{\rm c}$, $R_{\rm tot} \approx R_{\rm out}$ and the pressure drop will be concentrated at the inlet and outlet, with a constant pressure along the pore, as observed in previous MD works \cite{Suk2010}. 

Now these different estimates rely on the validity of the continuum hydrodynamics, which is expected to fail whenever the diameter of the pore is in the nanometer range \cite{Bocquet2010,Thomas2009}. Therefore we now explore the validity of the continuum predictions for entrance effects using MD simulations of flow across pores with sizes varying from nanometers down to a few angstr\"oms.

\section{Molecular dynamics simulations and finite-element calculations}

\subsection{Molecular dynamics}
\label{sec:md}

\begin{figure}
\includegraphics[width=\linewidth]{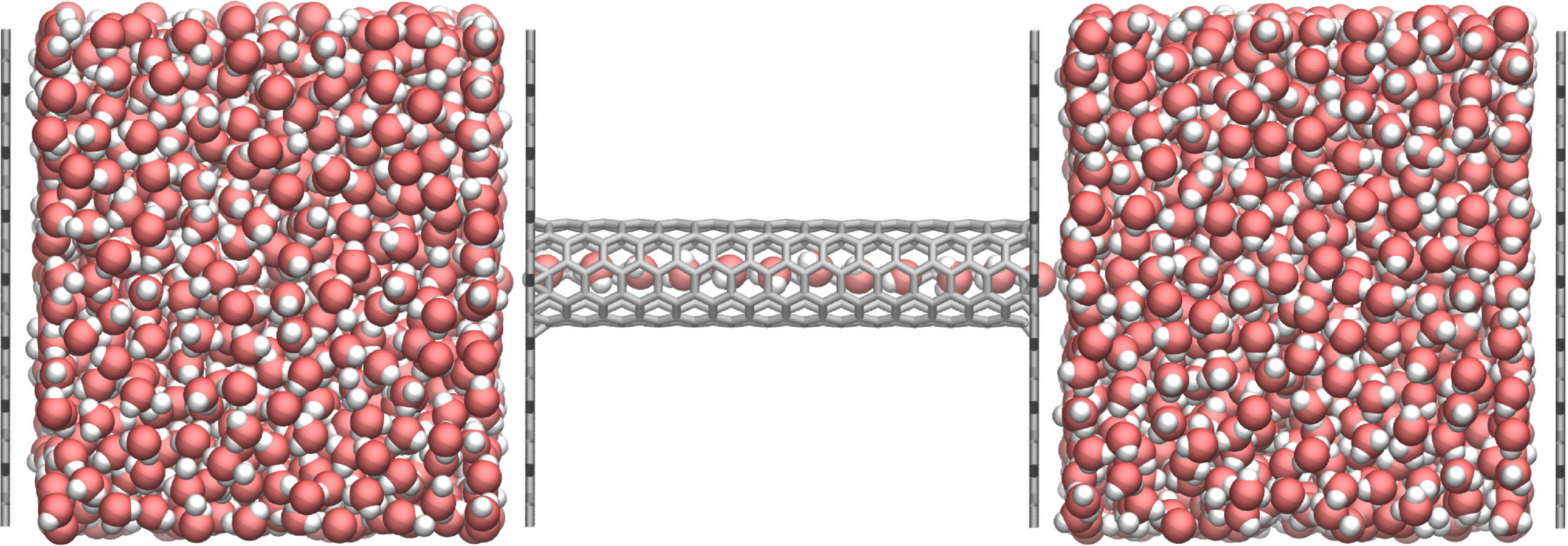}\\[2mm]
\includegraphics[width=\linewidth]{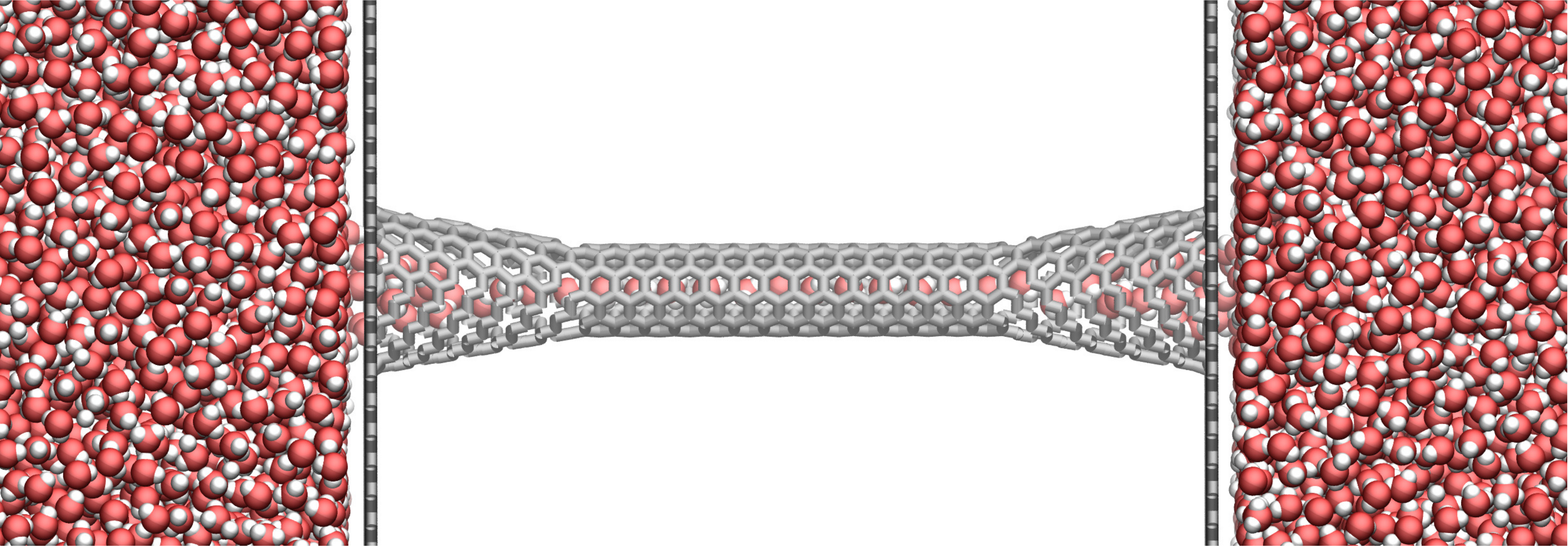}
    \caption{Snapshots of the systems considered. Up: Armchair (6,6) tube. Down: Biconical system [armchair (5,5) tube, $L_{\rm c}/a_{\rm eff}=20$ and $a_{{\rm c}}=3.38$\,\AA; see text for definition of the parameters]. Oxygen atoms are colored in red, hydrogen atoms in white, and carbon atoms in gray.}
\label{fig:illustration}
\end{figure}


In order to quantify entrance effects, we use geometries depicted in Fig.  \ref{fig:illustration}: two water reservoirs are separated by a membrane consisting of two parallel graphene sheets, and pierced with a nanopore. The pore can be a simple carbon nanotube (radius $a_{\rm c}$, defined by the center of carbon atoms, and length $L=10$ $a_{\rm c}$) or a biconical nanochannel, with a central nanotube and two conical entrances, made of graphene-like sheets. 
The inner tube radius $a_{\rm c}$ is varied from $25$\,\AA\ down to $3.5$\,\AA. We considered both armchair and zigzag nanotubes. 
The positions of the carbon atoms of the membrane are fixed (simulations with
flexible and fixed walls were shown to give similar results for the
statics and friction of confined liquids
\cite{Alexiadis2008,Thomas2009,Werder2003}). 
%
%
We use two pistons made of graphene sheets in order to maintain each reservoir at the desired pressure. 
\blue{Nanotubes are made by rolling a graphene sheet with respect to unit-cell.}
Periodic boundary conditions are imposed in all directions. The box size along the $z$ direction is taken sufficiently large to ensure that the periodic images do not interact. In order to avoid hydrodynamic interactions between the tube and its periodic images in the plane of the membrane, we choose a box with lateral dimensions ($x$,$y$) equal to 10 times $a_{\rm c}$. We also make sure that reservoirs are bigger than $10$ times $a_{\rm c}$  along $z$. 
There are 3.2 k water molecules and 2.7 k carbon atoms for the smallest system 
($a_c = 3.5$\,\AA),
and 800 k water molecules and 78 k carbon atoms for the biggest system 
($a_c = 25$\,\AA). 
Finite element calculations indicate that, in that configuration, the error due to finite size effects should be lower than $0.25\,\%$ (see following section). 

The Amber96 force field \cite{AMBER96} was used, with TIP3P \cite{Jorgensen1983}
water and water-carbon interaction modeled by a Lennard-Jones
potential between oxygen and carbon atoms, with parameters
$\varepsilon_{\mathrm{OC}} = 0.114$\,kcal/mol and $\sigma_{\mathrm{OC}} = 3.28$\,\AA. 
There is no need to define a potential between carbon atoms since they are fixed. 
%
%
The simulations were performed using LAMMPS \cite{LAMMPS}. Long-range Coulombic interactions were computed using the
particle-particle particle-mesh (PPPM) method \cite{Darden1993, Lu2003}, and water molecules
were held rigid using the SHAKE algorithm \cite{Ryckaert1977}. 
The equations of motion were solved using the velocity Verlet algorithm
with a timestep of $2$\,fs. 

Water molecules were kept at a temperature of 300\,K using a
dissipative particle dynamics (DPD) thermostat \cite{Groot1997}. 
This amounts to adding pairwise
interactions between atoms, with a dissipative force depending on the
relative velocity between each pair and a random force with a Gaussian
statistics. 
This method has the advantage of preserving hydrodynamics, even for complex 3-dimensional flows as the ones considered here. 
The amplitude of the dissipative term was carefully tuned to ensure that the liquid viscosity is negligibly affected by the thermostat 
(this thermostating method was extensively tested and compared to other approaches in a previous work \cite{Joly2011}).
%

\begin{figure}
\includegraphics[width=0.9\linewidth]{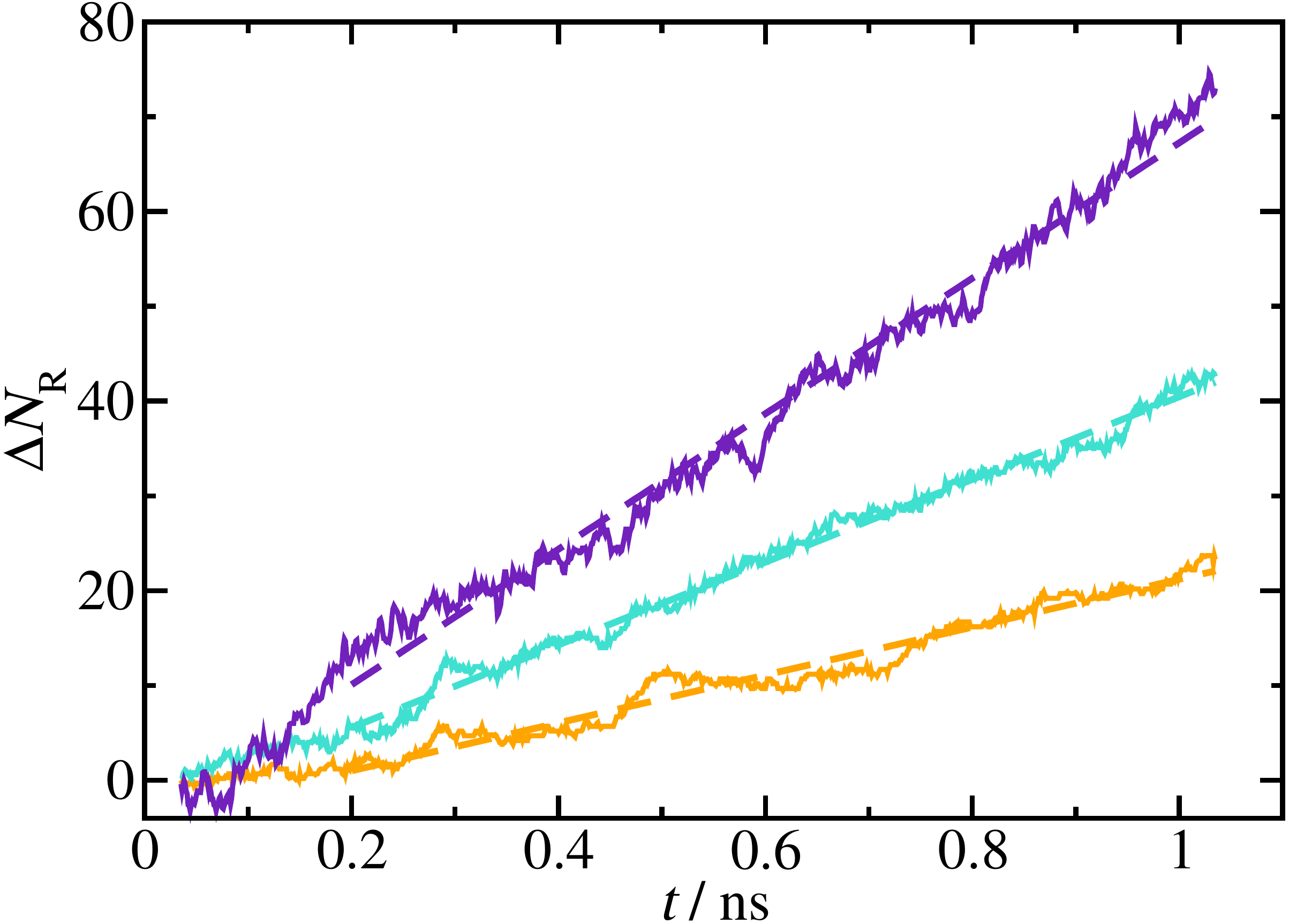}
    \caption{Number of crossing water molecules toward the right reservoir $\Delta N_R$  as a function of time, for different tube radii $a_{\rm c}$ 
($a_c=3.51$ \AA\ in yellow, 4.31 \AA\ in cyan, and 4.69 \AA\ in purple. Full lines represent the MD results, and dashed lines represent the linear fit used to compute the flow rate.}
\label{fig:method}
\end{figure}

Water molecules are initially disposed on a simple cubic lattice with
equilibrium density. Pressure differences are imposed using the reservoir pistons, and a steady-state flow quickly appears, within less than 200\,ps (see Fig. \ref{fig:method}). Note that this timescale matches the expected time for momentum diffusion in reservoirs, in agreement with the dominant role of entrance resistance.
%
%
%
We then measure the steady-state flow rate by counting the number of water molecules crossing the tube. This is shown in 
Fig. \ref{fig:method} for several tubes under a given pressure drop. The flux $Q$ is deduced from the linear fit of the  
time dependent variation of the number of crossing water molecules $\Delta N_R(t)$: $Q = M/(\rho \mathcal{N}_{\rm A}) \mathrm{d}\Delta N_R/\mathrm{d}t$, with $M$ and $\rho$ the molar mass and density of water, and $\mathcal{N}_{\rm A}$ the Avogadro constant. 
Finally, the hydrodynamic resistance is computed as $R=\Delta P/Q$. 
For each geometry we ran a number (up to 10) of independent simulations from different initial conditions, in order to estimate and reduce statistical uncertainties. The production times ranged from $\sim 1$\,ns for the largest pores, to $\sim 6$\,ns for the smallest ones.
%
%
%
Although the results presented in this article were obtained for a pressure drop of $1000\,\mathrm{atm} - 1\,\mathrm{atm} $, we emphasize that linearity between flux and pressure drop has been checked systematically in all our simulations.

\paragraph*{Slippage effects --}
Independently from the above calculation, we also calculated the slip length of  water on a graphene sheet. To this end, we performed simulations of Couette and Poiseuille flows of water confined between two graphene planes, in line with previous work  \cite{Huang2008,Sendner2009}. 
We measured a quite large slip length $b = 123 \pm 21$\,\AA. 
This means that  tubes with nanometer range diameters can be considered as perfectly slipping.
We also measured the water viscosity, and found it to be equal to $0.31 \pm 0.02$\,mPa.s, in good agreement with the expected value of $0.31$\,mPa.s for the TIP3P model at 300\,K \cite{Wu2006}.

\subsection{Finite element calculations}
\label{sec:fem}

\begin{figure}
\includegraphics[width=0.8\linewidth]{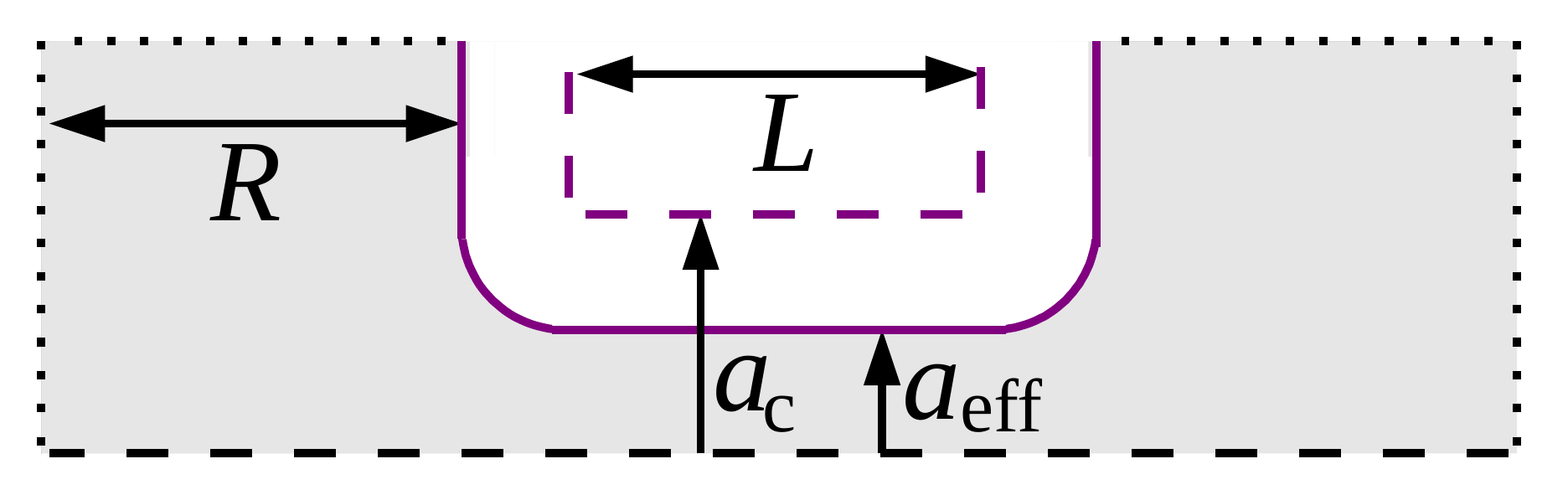}
\includegraphics[width=0.8\linewidth]{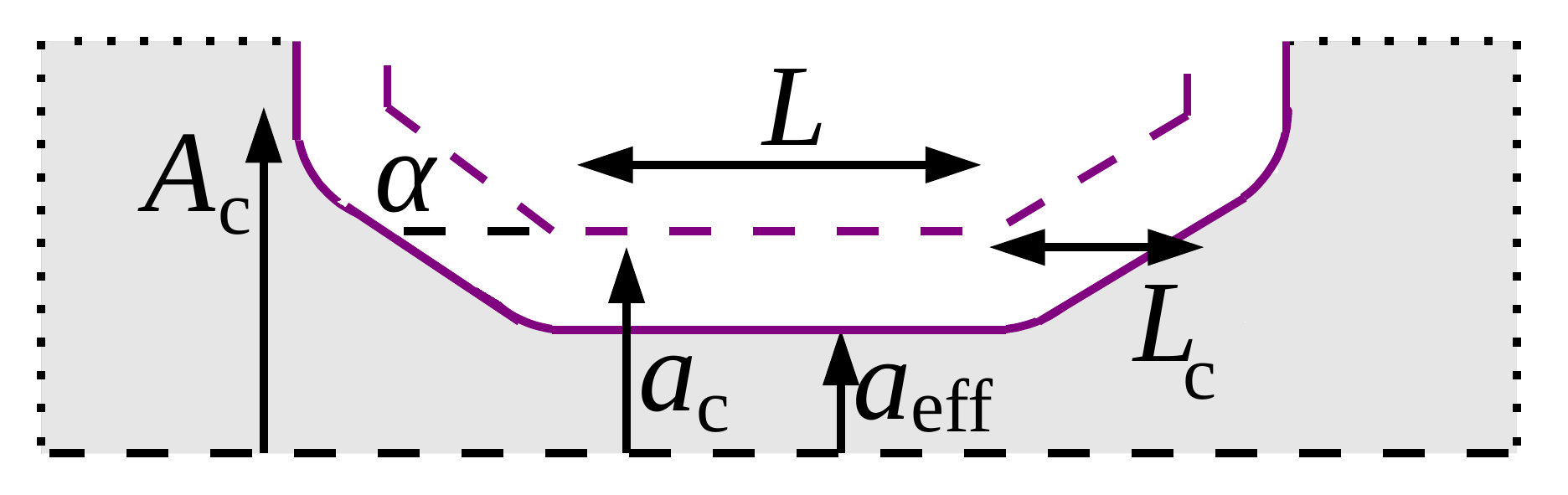}
    \caption{Cylindrical and hourglass geometries used for the FE calculations. Two reservoirs of size $L_{\rm r}$ are separated by a membrane containing a channel of radius $a_{\rm eff}$ and length $L$. The black dashed line corresponds to an axisymmetric condition. The full line represent the liquid/solid interface, we use perfect slip BC along it. We impose the pressure on the dotted lines. $A_c$ is the opening radius and $\alpha$ the cones angle.}
\label{fig:systemcomsol}
\end{figure}

In order to compare MD results with continuum predictions, we perform FE calculations using  the commercial software COMSOL.
We solve the Stokes equation $\eta \Delta \vec{v} = \vec{\nabla} p$ in a 2D-axisymmetric geometry. For the cylindrical pore, the system is composed by a tube of radius $a$, and length $L=10 a$. Similar calculations are performed with the hourglass shape \cite{Gravelle2013}, see Fig. \ref{fig:systemcomsol}.
The pore is connected to two reservoirs of characteristic size $L_{\rm r}$. Far from the pore, we impose a difference of pressure $\Delta p$ between the two reservoirs and then measure the water flow across the tube, which gives us the hydrodynamical resistance $R = \Delta p / Q$. Along membrane and tube walls we use perfect slip boundary conditions. 
Note that a small but finite friction may be expected 
at the water/carbon surface. However using typical slip lengths for the considered systems, as reported in the previous section, 
we checked that the corresponding effect on the hydrodynamic resistance is negligible as compared to entrance effects: typically, 
 the finite slip length affects the total resistance by $\sim 0.2\,\%$ for $a_c=3$\,\AA\  and less than $10\,\%$ for $a_c=22$\,\AA, and thus does not bring significant changes.

Note that we use two different values of the radius $a$: $a_c$, which is the distance between the centers of carbon atoms, and $a_{\rm eff}$, which is the hydrodynamic radius of the pore, see Fig. \ref{fig:systemcomsol}. Typically one expects the hydrodynamic boundary condition (BC) to apply within the fluid at a distance of $\sim \sigma_{OC}$, {\it i.e.} one molecular size from the wall \cite{Bocquet1994}. Accordingly we expect for the radius
$a_{\rm eff} \approx a_c - \sigma_{OC}$ (this will indeed be in agreement with our results). However in the following we will merely use $a_{\rm eff}$ as a free parameter, in order to identify the best agreement between the MD and FE results. For consistency, we introduce chamfers at the wall corners, with a radius of curvature $a_c-a_{\rm eff}$.


We also used FE calculations to estimate the minimal simulation box size needed to keep finite size effects to an acceptable level. Indeed, far from the entrances, the flow is identical to the one from a point source on a plane wall, implying that the liquid velocity decreases algebraically and size effects due to the finite size of the reservoirs in the MD simulations could therefore affect  the flow rate versus pressure drop measurements. We checked that a characteristic size of reservoir $L_{\rm r} \sim 10 a_{\rm eff}$ ($a_{\rm eff}$ the pore radius) ensures an error of ca. $0.25\,\%$.




\section{Entrance effects at the molecular scale: the cylindrical pore geometry\label{sec:cyl}}


\begin{figure}
\includegraphics[width=0.9\linewidth]{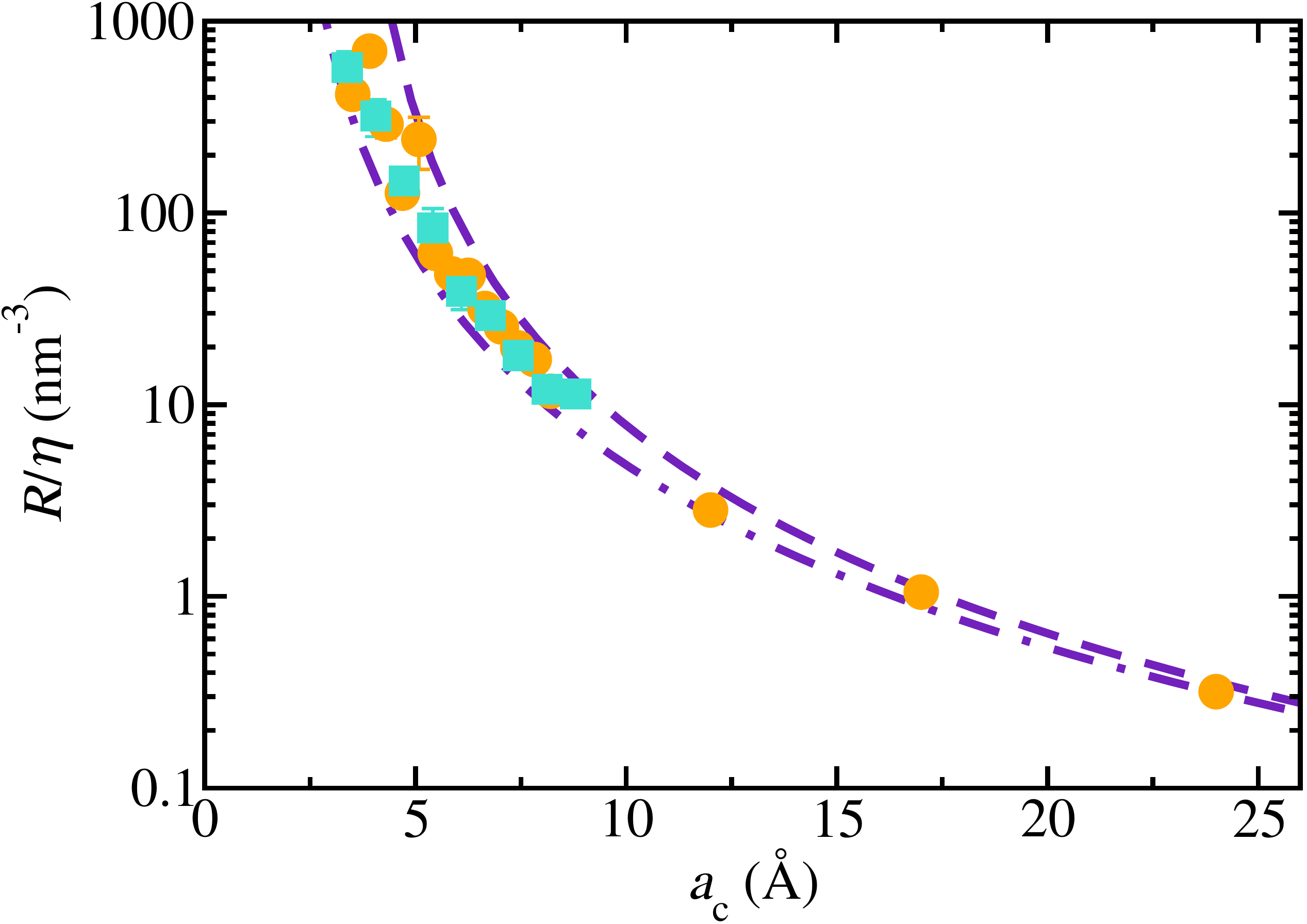}
    \caption{Resistance of a single nanotube $R = \Delta P / Q$, normalized by the bulk liquid viscosity $\eta$, as a function of the radius $a_{{\rm c}}$ of the tube. Circles represent MD results with armchair tubes, squares represent MD results with zigzag tubes. Lines are hydrodynamic predictions using FE calculations, for $a_c-a_{\rm eff}=3.2$\,\AA\ (dashed dotted line) and $a_c-a_{\rm eff}=1.7$\,\AA\ (dashed line).}
\label{fig:R_vs_a}
\end{figure}

Let us first explore the flow resistance across cylinders of finite length. 
%
Following the procedure described in section \ref{sec:md}, 
the hydrodynamic resistance $R=\Delta P/Q$ is extracted for the various tube diameters using MD simulations. 
We plot in Fig. \ref{fig:R_vs_a} the resistance normalized by the bulk viscosity, $R / \eta$, as a function of the tube radius $a_{{\rm c}}$. 
%
These data are compared with 
continuum predictions obtained using FE calculations in the same geometry.
%
%
We used two different values for the hydrodynamic pore radius: one where the perfect slip hydrodynamic BC is located right at the surface of carbon atoms, {\it i.e.} 
$a_c-a_{\rm eff}=\sigma_{\rm c}/2 \approx 1.7$\,\AA;  and the other where the perfect slip hydrodynamic BC is located at the first layer of water along the wall $a_c-a_{\rm eff} \approx \sigma_{OC} \approx 3.2$\,\AA, in line with results from Ref. \cite{Bocquet1994,Huang2008}.
%
%
As shown by Figure \ref{fig:R_vs_a}, we obtain a very good agreement between the MD results and the continuum hydrodynamic predictions obtained by FE. In line with previously quotted expectations, we found that a reasonable choice for the value of the hydrodynamic radius is $a_{\rm eff}\simeq a_c-2.5$ \AA. 
Note that nanotube's chirality has no significant influence on entrance effects. 

It is particularly interesting to observe that the hydrodynamic prediction is valid for nanotubes with effective radii well below one nanometer, even when single file transport occurs. In such case, we expect a full breakdown of hydrodynamics. 
However, entrance effects originate mainly from the bending of the streamlines {\it outside} the tube, which occurs in the bulk on length scales much larger than the tube radius. This certainly explains the robustness of hydrodynamics to predict entrance effects.

 

\section{Entrance effects at the molecular scale: hourglass nano-pores}

\subsection{Molecular flows through the hourglass geometry}

We now turn to the hourglass geometry. In our previous work \cite{Gravelle2013},
we showed on the basis of continuum hydrodynamic calculations that the hourglass shape, inspired by aquaporins, lead to a large increase of the overall channel permeability as compared to the cylindrical geometry. 
Here we explore how this conclusion can be extended to the case where the liquid in the middle part of the channel is subjected to single file transport, as occurs {\it e.g.} in aquaporin channels.

As discussed above, we build up a biconical nanochannel using graphene-like walls, see Fig. \ref{fig:illustration}. The cones exhibit an angle $\alpha$, which is varied between  $0$ and $20\,^\circ$. 
\blue{Conical entrance are made alike by rolling graphene sheet, thus leaving now a structural defect line. Note that the overall pore permeability is not affected by this defect line as tested using alternative methods to generate the cones.} 
We choose a configuration in which there is single-file flow inside the central part, as found in aquaporins. Looking for possible departures from continuum hydrodynamics, we consider the most extreme case, with the smallest radius that allows water molecules to fill the tube ($a_{\rm c} = 3.38$\,\AA). 
Inspired by aquaporins, we choose cones' lengths $L_{\rm c}  \approx 20 \times a_{\rm eff} = 17$\,\AA.
We use reservoirs with a size at least ten times the opening radius, 
$A_{\rm c}$ (with $A_{\rm c}=a_{\rm eff} + L_{\rm c} \tan\alpha$), which is sufficient to reduce finite size effects down to a negligible level, as estimated by FE calculations (see section \ref{sec:fem}). 


%

\begin{figure}
\includegraphics[width=0.9\linewidth]{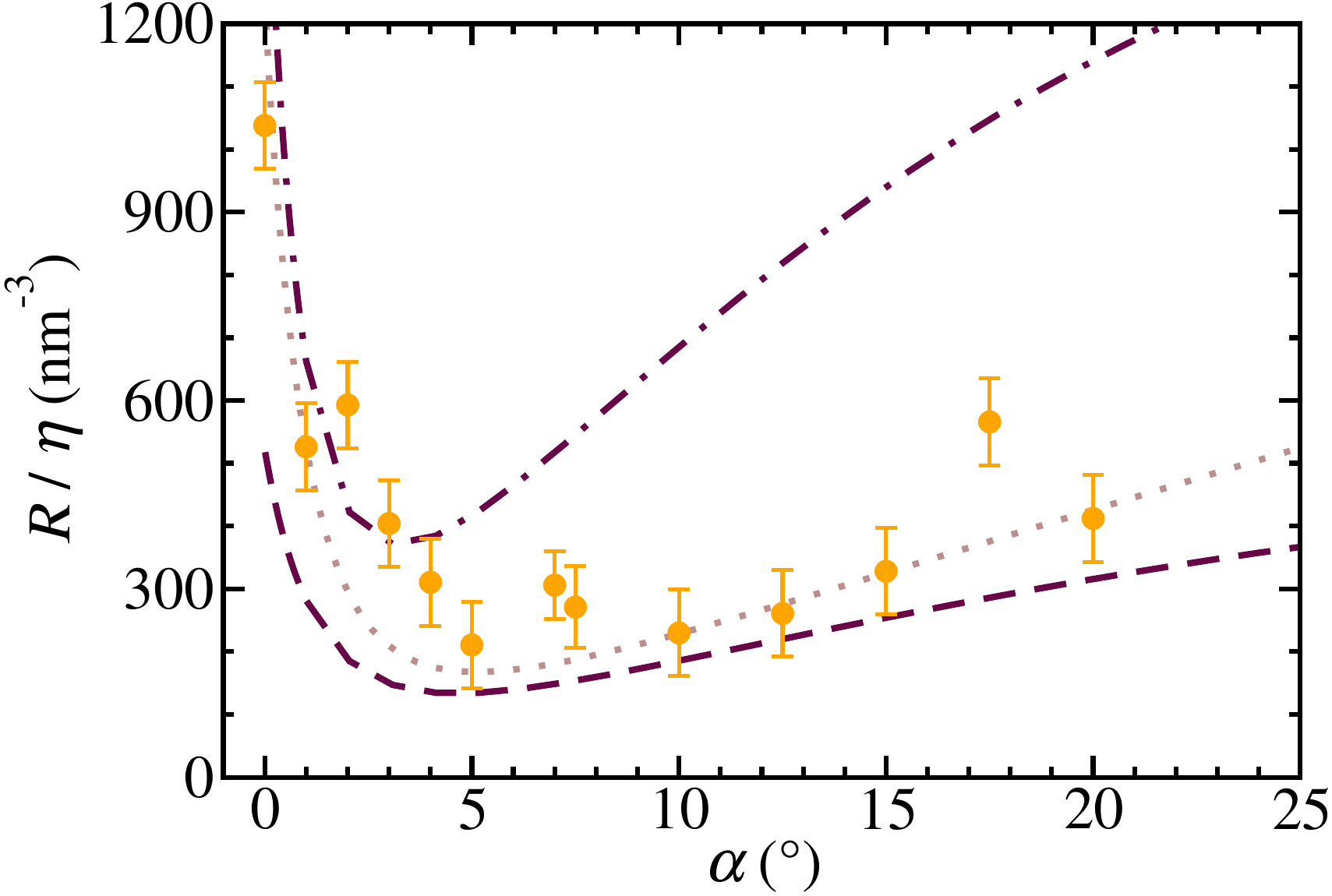}
    \caption{MD results for the hydrodynamic resistance $R=\Delta P / Q$ of an hourglass channel (normalized by the bulk viscosity $\eta$) as a function of the angle of the biconical vestibules. Circles represent MD results. The dotted line represents the prediction of Eq. (\ref{eq:theory}), with $a_{\rm eff}=1$\,\AA. The lines show the FE calculations with two different hydrodynamic radii: $a_{\rm eff}=1.08$\,\AA\ (dashed line), $a_{\rm eff}=0.68$\,\AA\  (dashed-dotted line).
}
\label{fig:bicone}
\end{figure}

The MD results are shown in Fig. \ref{fig:bicone}. We find  that the hydrodynamic resistance is again minimized in the hourglass geometry for a cone angle $\alpha \sim 5\,^\circ$, \emph{even if the transport in the center part of channel is single-file}. The minimal resistance is ca. 5 times smaller than the one of a tube with a straight entrance ($\alpha = 0$). 
Furthermore, 
as demonstrated in Fig. \ref{fig:bicone}, the continuum hydrodynamic predictions exhibit a good agreement with the results of the MD simulations  for the hydrodynamic resistance. In particular, we highlight the critical importance of the position of the hydrodynamic BC by showing two different cases of hydrodynamic radius for the inner part of the pore. 

We finally compare the results with our theoretical prediction for the hourglass resistance obtained in Ref. \cite{Gravelle2013}, which writes
\begin{equation}
\frac{R}{ \eta}  = \frac{C}{a_{\rm eff}^3} \left[ \sin\alpha + \left(1 + \frac{L_c}{ a_{\rm eff}} \tan \alpha \right)^{-3} \right]
\label{eq:theory}
\end{equation}
We used the values $L_c/a_{\rm eff}=20$, and the parameter $C=1.15$ obtained from previous FE calculations in the case of an aspect ratio $a_{\rm eff}/(a_{\rm c}-a_{\rm eff})$ = 0.33 (see Appendix).  
The effective radius of the biconical channel was fixed to $a_{\rm eff} = 1$\,\AA$ = a_c - 2.4$\,\AA\ in order to get best agreement. With this value, Eq. (\ref{eq:theory}) reproduces quite well the MD results (see Fig. \ref{fig:bicone}). Note that MD results are best fitted with FE calculations for the same value of $a_{\rm eff}$ (not shown for clarity). Therefore, the ``hydrodynamic size'' of the wall atoms obtained with the biconical geometry, $a_c - a_{\rm eff} = 2.4$\,\AA, is quite close to the one obtained with a cylindrical pore in previous section ($a_c - a_{\rm eff} = 2.5$\,\AA). 
%
Additionally, one may notice that a number of MD results are slightly outlying the main tendency, beyond the error bar. We conjecture that these may be attributed to discrete effects which may occur preferably for specific values of the cone angles. 

\subsection{Comparison with pierced graphene}

Finally, we compare the efficiency in terms of transport, of hourglass shape nanopores versus pierced graphene. 
Pierced graphene was discussed lately to represent a very efficient geometry for desalination purpose \cite{Cohen-Tanugi2012,Suk2013,O'Hern2014}, in particular due to its large permeability combined with a excellent rejection ability.
The large permeability of pierced graphene is due to the molecular thickness of the graphene sheet: in this case, the transport is fully controlled by the entrance effect, and belongs to the Sampson's class of problem. It is therefore interesting to compare the performance of this geometry to the hourglass one which was precisely found to reduce entrance effects.

\begin{figure}
\includegraphics[width=0.9\linewidth]{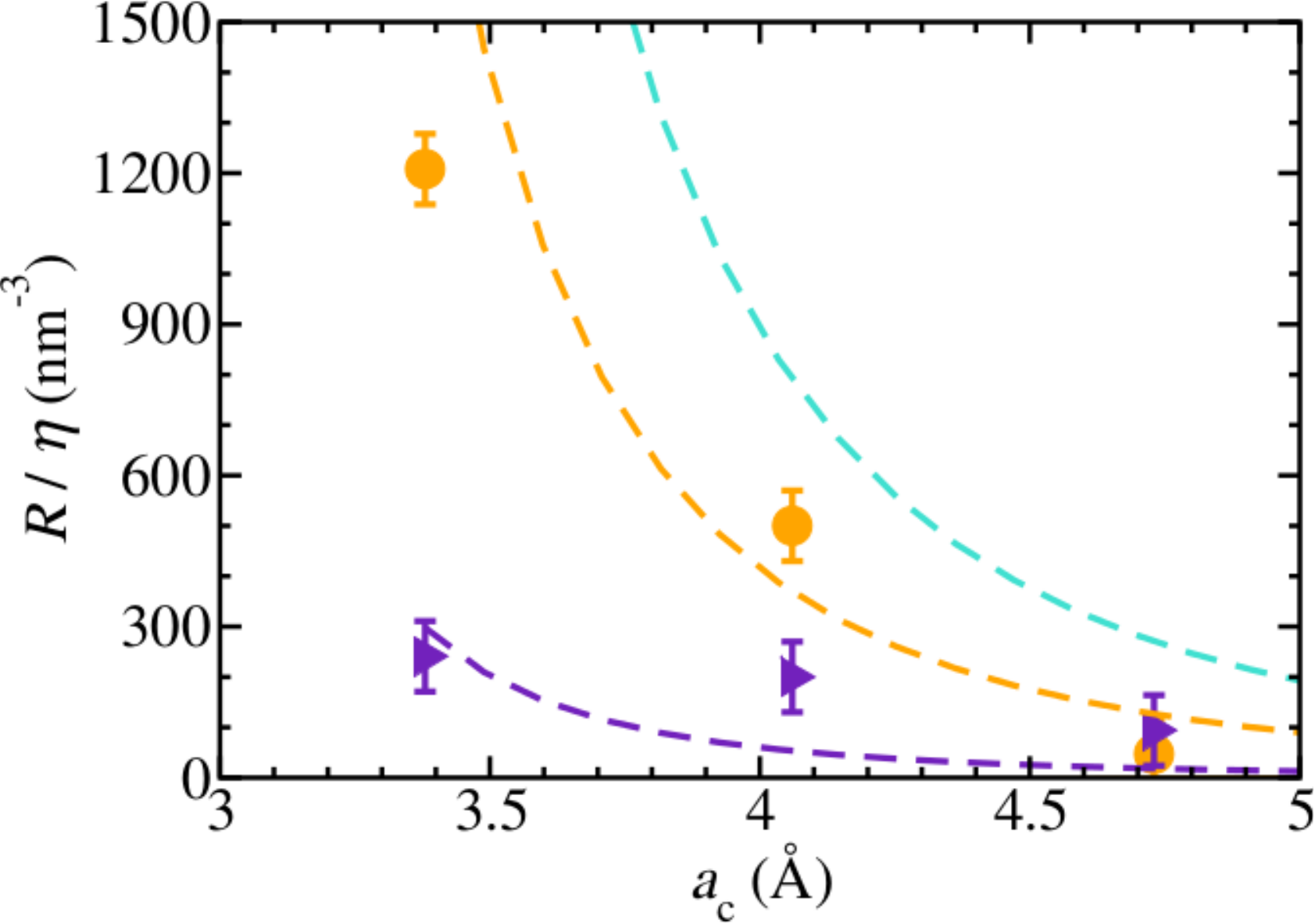}
    \caption{Comparison of hydrodynamic resisttance of hourglass, biconical systems (purple triangles) and graphene sheets pierced with a single hole (orange circles). Dashed lines are continuum hydrodynamics predictions (see text). The cyan dashed line is the Sampson formula for a hole in a infinitely thin membrane.}
\label{fig:aqpgraphene}
\end{figure}

To this end we have performed MD simulations of transport across nanopores drilled in a graphene sheet with various pore sizes, and measured the corresponding hydrodynamic resistance. This is compared to the hydrodynamic resistance of an hourglass shape nanopore with an inner channel having the same radius. {We use hourglass systems with three different inner pore radii, respectively $a_c= 3.38$, $4.06$ and $4.73$\,\AA. We keep the cone lengths ratio $L_c/a_c$ equal to 20 and the angle $\alpha$ equal to 5$\,^\circ$. The radius of the hole $a_c$ in the graphene sheet is  equal to $3.38$, $4.06$ and $4.73$\,\AA.}
Results are shown in Fig. \ref{fig:aqpgraphene}. 
%
%
As can be seen in this figure, the hourglass nanopore has a hydrodynamic resistance which is systematically smaller
than the graphene for the same inner pore diameter.
%
As a comparison, we plot the result of the classic Sampson formula, Eq. (\ref{sampson}), with a hydrodynamic radius equal to $a_{\rm eff}= a_c-2.5$\,\AA\ (see above) and a coefficient $C=3$. 
As seen in Fig. \ref{fig:aqpgraphene}, this overestimates the MD results for the hydrodynamic resistance of porous graphene. In an attempt to improve this result, we take into account at the continuum level the finite thickness of the graphene membrane, as compared to the pore radius (see appendix for details). This leads to a reduction of the coefficient $C$ in the Sampson formula. 
%
%
The value of $C$ depends on the ratio $a_{\rm eff}/(a_{\rm c}-a_{\rm eff})$, and in the present conditions it varies between 1.15 and 1.6 for the considered radii. We choose $C\simeq 1.3$ in Eq.(\ref{sampson}) as a compromise, showing a good agreement with the MD results.
For the hourglass geometry, we compare the MD results with the prediction of Eq. (\ref{eq:theory}), using the same coefficient $C=1.3$.
This shows again a good agreement with the MD results.

Altogether these results show that tuning the geometry of nanopores allow to strongly optimize water transport through membranes.
The hourglass shape outperforms both nanotubes and pierced, molecular thick, graphene. 

\section{Conclusion}

In this paper, we investigated the fate of hydrodynamic entrance effects in the limit of molecular scale nanopores, down to the single-file regime. Strikingly, we found an extremely good agreement between MD simulations and continuum hydrodynamics predictions of entrance resistance obtained from FE calculations. For straight nanopores, comparison between MD and FE approaches incorporate the position of the solid-liquid interface as an important parameter, which therefore could be located using such an approach. In addition, the agreement between molecular scale transport and continuum hydrodynamics extends to more complex geometry of nanopores, inspired by the hourglass shape of aquaporin biological nanochannels. 

This hourglass shape proves to substantially improve the water transportation efficiency down to single-file regime, with an optimum for shallow opening angles $\sim5^\circ$, all in line with continuum hydrodynamics predictions. Compared with the promising system made of a circular nanopore drilled in graphene sheets, or straight carbon nanotubes, the hourglass shape pore appears far more efficient, illustrating the dominant role of entrance dissipation in all these systems. Overall, this stresses the necessity for finely tune the geometry of nanopores for strongly optimizing water transport across membranes, a task for which simple continuum approaches can be astonishingly reliable.

%
%
%


\begin{acknowledgments}
The authors acknowledge financial support from ERC-AG project Micromegas. Most of those computations were performed at the P\^ole Scientifique de Mod\'elisation Num\'erique (PSMN) at the \'Ecole Normale Sup\'erieure (ENS) in Lyon. We are grateful for the computing resources on JADE (CINES, French National HPC) obtained through Projects c20132a7167 and c20142b7167.
\end{acknowledgments}

\appendix

\section*{Appendix: Sampson formula and membrane size effect}

\begin{figure}
\includegraphics[width=0.8\linewidth]{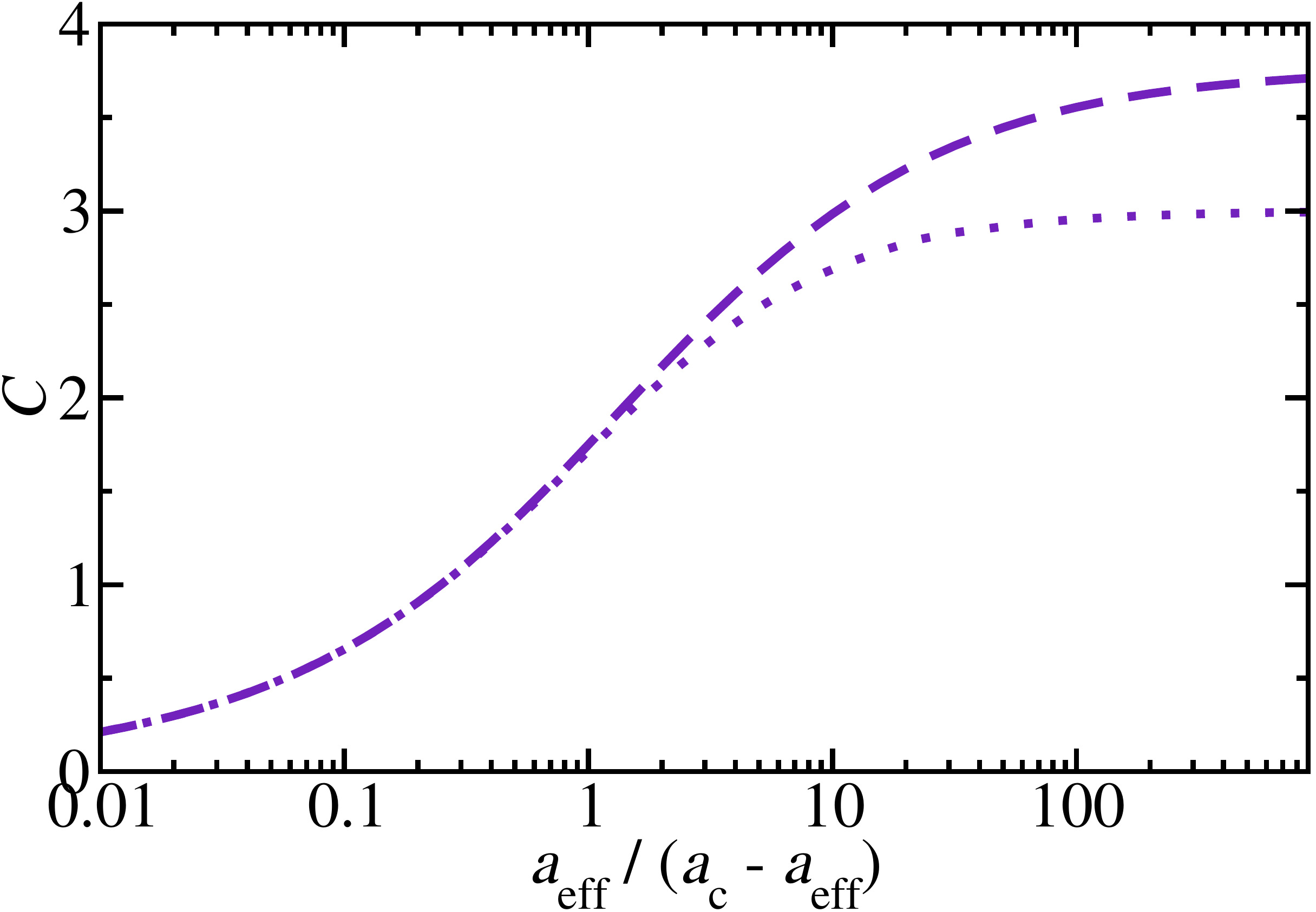}
    \caption{FE testing of the Sampson formula with a membrane of finite thickness: Measured Sampson coefficient as a function of the ratio effective radius $a_{\rm eff}$ over chamfer radius $a_{\rm c}-a_{{\rm eff}}$. (long dashed line) nanotube limit $L/a_{{\rm eff}}=5$; (dashed line) atomically thin membrane nanopore limit $L=0$.}
\label{fig:finitesize}
\end{figure}

Carbon nanotubes are made of atoms with a finite radius, and the graphene membrane has therefore a finite molecular width which could compare to the diameter of the pore, for nanometric pores.   At the level of continuum description, this finite aspect ratio is expected to modify the prefactor $C$ in the Sampson equation,  Eq. (\ref{sampson}). We estimate here the modified coefficient $C$ using FE calculations.
%
%
%
The effective radius of carbon atoms (defining the hydrodynamic radius of the pore) is equal to $a_{\rm c}-a_{\rm eff}$, see figure \ref{fig:systemcomsol} in the main text. 
We explored a large range of radius $a_{{\rm eff}}$ for fixed $a_{\rm c}-a_{{\rm eff}}$. Figure \ref{fig:finitesize} shows the effect of the aspect ratio on the value of the $C$ coefficient, for 2 different membrane thickness. As we can see, for a ratio $a_{\rm eff}/(a_{\rm c}-a_{\rm eff}) \approx 1$, which correspond to a (5,5) armchair tube, $C$ is around 1.5 instead of 3.75, corresponding to a $60\,\%$ improvement of the pore permeability. Note that this upper value of $C=3.75$ corresponds to the thick membrane limit achieved for our hourglass pore and that Sampson's prediction $C=3$ is recovered for vanishingly thin membranes as can be seen in figure \ref{fig:finitesize} for $L=0$. Surprisingly, the frictionless inner pore section increases the global, entrance-associated, dissipation as discussed in our recent work \cite{Gravelle2013}.

\end{document}